# Multi-View Substructure Learning for Drug-Drug Interaction Prediction

Zimeng Li[1,2†], Shichao Zhu[3,4,2†], Bin Shao[2*], Tie-Yan Liu[2], Xiangxiang Zeng[1*] and Tong Wang[2*]

[1]College of Information Science and Engineering, Hunan University, Changsha, 410086, China.
[2]Microsoft Research Asia, Beijing, 10080, China.
[3]School of Cyber Security, University of Chinese Academy of Sciences, Beijing, 100049, China.
[4]Institute of Information Engineering, Chinese Academy of Sciences, Beijing, 100093, China.

*Corresponding author(s). E-mail(s): binshao@microsoft.com; xzeng@hnu.edu.cn; watong@microsoft.com;
[†]These authors contributed equally to this work.

## Abstract

Drug-drug interaction (DDI) prediction provides a drug combination strategy for systemically effective treatment. Previous studies usually model drug information constrained on a single view such as the drug itself, leading to incomplete and noisy information, which limits the accuracy of DDI prediction. In this work, we propose a novel multi-view drug substructure network for DDI prediction ("MSN-DDI"), which learns chemical substructures from both the representations of the single drug ("intra-view") and the drug pair ("inter-view") simultaneously and utilizes the substructures to update the drug representation iteratively. Comprehensive evaluations demonstrate that MSN-DDI has almost solved DDI prediction for existing drugs by achieving a relatively improved accuracy of 19.32% and an over 99% accuracy under the transductive setting. More importantly, MSN-DDI exhibits better generalization ability to unseen drugs with a relatively improved accuracy of 7.07% under more challenging inductive scenarios. Finally, MSN-DDI improves prediction performance for real-world DDI applications to new drugs.

# Introduction

Drug combinations can provide therapeutic benefits but also increase the risk of adverse side effects, caused by the physicochemical incompatibility of the drugs [1, 2, 3]. The identification of drug–drug interactions (DDIs) remains a challenging task considering that the huge number of drug combinations lead to the pharmaceutical research and clinical trials highly expensive and inefficient, even with high-throughput methods. There lots of computational methods for prediction of side effects caused by DDIs have emerged, which have proven to be an effective and alternative way to alleviate the challenge [4, 5, 6, 7, 8]. Most of these methods follow the assumption that drugs with similar features are more likely to have similar interactions. In order to make full use of the raw features of drugs, i.e., drug structures, chemical properties and molecular fingerprints, recent works mainly focus on utilizing the powerful feature extraction ability of deep neural networks [9, 10, 11, 12]. As a drug can be represented as a graph based on its molecular structure, graph neural networks (GNNs) have shown the impressive representation learning ability of drug molecules. Existing GNN-based methods for DDI [13, 14, 15] usually take the advantage of GNN's topological and semantic representation capabilities to model the drug itself, and then learn the representation of drug pairs based on the respective representation of each drug. Finally, the representations of drugs or drug pairs are used for final DDI prediction.

Considering that a drug can be simply divided into several functional groups or chemical substructures which jointly lead to the overall pharmacological properties [16], some studies were motivated to refine drugs into substructures for DDI prediction [17, 18, 19, 20]. Existing works can be roughly classified into two categories: implicit and explicit manners, depending on how the substructure is used. The implicit manner usually takes substructure features as inputs of the model, which doesn't explicitly learn a specific substructure through the neural network[17, 18]. As a contrast, the other approach, including SSI-DDI [19], GMPNN-CS [20] and so on, extracts the respective substructures of a pair of drugs in drug representation learning stage and predict DDI effect by identifying pairwise interactions between two drugs' substructures in the final readout module, leading to an improvement in performance over previous methods. However, the extracted substructures of a drug pair are only combined and used in the readout module for final DDI prediction instead of playing a direct role in the drug representation learning.

In most DDI prediction algorithms, drug representation learning is a single-view process in the message passing module that only encodes information from the drug itself, which may hinder accuracy improvement of DDI prediction. There are some interests that try to adopt multi-view representation learning into DDI prediction, such as MHCADDI [21], GoGNN [22] and MIRACLE [23]. For example, MHCADDI [21] considers external message passing mechanism between drugs' structures to integrate joint drug–drug information during the representation learning phase for individual drugs. GoGNN [22] leverages the dual attention mechanism to capture the information from both entity graphs and structured-entity interaction graph, hierarchically. However, multi-view learning in these methods only serves to learn better drug representations while it could be further employed in the readout module for final DDI prediction.

Therefore, unlike the above methods, by employing the advantages of both substructures and multi-views, we propose a novel multi-view substructure learning for DDI prediction (termed as "MSN-DDI"), which learns substructures from intra-view and inter-view simultaneously, without depending on additional domain knowledge. This makes the model equally applicable to inductive settings where only the chemical structure of the drug itself is accessible. MSN-DDI consists of the following main components, including repetitive multi-view substructure extraction blocks as the encoders (MSN encoders) to model different orders of neighboring information, layer-wise substructure pooling layers as the substructure extraction module to learn substructures from different perspectives and the self-attention scoring function as the MSN decoder for final DDI prediction. Specifically, we regard the drug representations as the intra-view and drug pair interactions as the inter-view and thus define graph attention network layers respectively to learn two sets of substructures corresponding to each view. The two sets of substructures are further used to update node representations for the next MSN encoder block. Our comprehensive evaluation on DrugBank and Twosides datasets demonstrated that MSN-DDI has achieved 19.32% relative accuracy improvement on the transductive setting and 7.07% relative improvement to unseen drugs on the inductive setting. Furthermore, the AUC reaches 99.47% on DrugBank and 99.90% on Twosides for the transductive setting respectively, which indicates MSN-DDI almost has solved the DDI prediction task for existing drugs. In addition, MSN-DDI exhibits the usefulness of DDI prediction to new approved drugs and could also show some clues to interpret the DDI effect to interactions among substructures of the drug pair. All these results suggest that MSN-DDI can act as a useful tool for DDI prediction and thus greatly facilitate the drug design and discovery process.

# Results

## Overview of MSN-DDI architecture

Inspired by recent advances in substructure and multi-view representation learning, our approach learns the drug representation and drug-drug interaction from inter-view and intra-view simultaneously which achieves better results compared to state-of-the-arts on DDI prediction. As shown in Figure 1, MSN-DDI consists of the following components:

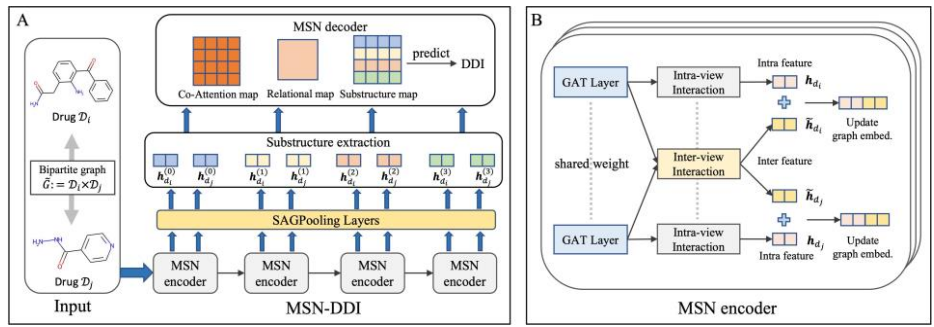

**Fig. 1** Schematic of the MSN-DDI architecture. (A) The framework of MSN-DDI.The input drug pair is encoded by a bipartite graph followed by a series of repetitive MSN encoder blocks. For each block, substructures are extracted by the substructure module, in which substructure specific embeddings $\boldsymbol{h}_{d_i}^{(l)}$, $\boldsymbol{h}_{d_j}^{(l)}$ are summed up based on SAGPooling layer.

Finally, all substructures are fed into MSN decoder that is defined as a co-attention scoring function for a given triplet for DDI prediction. (B) MSN encoder. The drugs are firstly encoded by a shared GAT layer, and then embedded by intra-view interaction and inter-view interaction modules through two dedicated GAT layers. The inter-view and intra-view information are then aggregated to update node representations for the next MSN encoder.

- MSN encoder: Following a bipartite graph to encode the input features of a drug pair, a series of repetitive MSN encoder blocks capture the interactions within drug (intra-view) and across drug boundaries (interview) simultaneously. For each drug, two dedicated GATs are designed following a shared GAT layer in each block to learn atom-level representations from both views. The inter-view and intra-view information are then aggregated to update node representations for the next MSN encoder.

- Substructure extraction module: Followed by a MSN encoder block, a self-attention graph pooling layer is designed to learn and extract substructure representations for both drugs. Since a series of encoders capture different orders of neighbouring information, the substructures following these encoders are extracted from different perspectives.

- MSN decoder: a co-attention scoring function, to predict the probability of the triplet ($d_i$, $r$, $d_j$), where $d_i$ and $d_j$ stand for the drug pair and $r$ stand for a type of drug pair interaction. In this component, each pair of drug's substructures is integrated by how much important or relevant it is to the final DDI prediction.

## Performance evaluation on transductive Setting for existing drugs

We conduct experiments on two standard benchmarks: DrugBank and Twosides, to evaluate the performance of our method MSN-DDI. The statistics of the datasets are summarized in Table S1. For both datasets, each drug is associated with its SMILES string [24], and its molecular graphical representation is converted from SMILES using the python library RDKit [25], which contains 55-dimensional initial chemical features for each atom, such as atomic symbols and degrees of the atoms.

**Table 1** Performance evaluation between MSN-DDI and baselines for the transductive setting on DrugBank and Twosides datasets. The highest value in each column is shown in bold. For performance improvement over the second-best approach, a relative improvement percentage is shown in the bracket.

| Method | DrugBank | | | | Twosides | | | |
|---|---|---|---|---|---|---|---|---|
| | ACC(%) | AUC(%) | AP(%) | F1(%) | ACC(%) | AUC(%) | AP(%) | F1(%) |
| MR-GNN | 96.04±0.05 | 98.87±0.04 | 98.57±0.06 | 96.10±0.05 | 76.23±0.23 | 85.00±0.22 | 84.32±0.35 | 77.88±0.35 |
| MHCADDI | 83.80±0.27 | 91.16±0.31 | 89.26±0.37 | 85.06±0.31 | - | 88.20 | - | - |
| SSI-DDI | 96.33±0.09 | 98.95±0.08 | 98.57±0.14 | 96.38±0.09 | 78.20±0.14 | 85.85±0.13 | 82.71±0.14 | 79.81±0.16 |
| GAT-DDI | 89.81±1.00 | 95.21±0.70 | 93.56±0.90 | 90.18±0.74 | 50.00 | 50.00 | 50.00 | - |
| GMPNN-CS | 95.30±0.05 | 98.46±0.01 | 97.94±0.02 | 95.39±0.05 | 82.83±0.14 | 90.07±0.12 | 87.24±0.12 | 84.08±0.14 |
| MSN-DDI | **96.94±0.02** | **99.47±0.01** | **99.37±0.02** | **96.93±0.02** | **98.83±0.04** | **99.90±0.01** | **99.89±0.01** | **98.83±0.04** |
| Improvement | +0.61(0.63%) | +0.52(0.53%) | +0.80(0.81%) | +0.55(0.57%) | +16.00(19.32%) | +9.83(10.91%) | +12.65(14.5%) | +14.75(17.54%) |

Similar to previous studies, we evaluate performance on two settings, transductive and inductive. For the transductive setting, drugs in the test sets also exist in the training set while the inductive setting contains drugs fully or partially not existing in the training set to examines the model generalization ability to new drugs. Following the same setting in the related work [20], for the transductive setting evaluation on DrugBank and Twosides, we perform three randomized folds with the same data split ratio of training:validation:test = 6:2:2, where the stratified split on both datasets performed on entire DDI triplets, including drugs and interactions. Furthermore, to make a fair comparison, MSN-DDI also adopts the standard deep learning experiment process and the same datasets for positive samples and negative samples with all baselines. Experiment results are reported with the means and standard deviations of the following six metrics across the three folds: the accuracy (ACC), the area under the receiver operating characteristic (AUC), the average precision (AP), the F1 score. The detailed definitions of each metric are given as follows.

- the accuracy (ACC): is defined as the number of correct predictions divided by the number of total predictions.
- the area under the receiver operating characteristic (AUC): is equal to the probability that a classifier will rank a randomly chosen positive instance higher than a randomly chosen negative one.
- the average precision (AP): is calculated by taking the mean average precision over all classes.
- the F1 score: is the harmonic mean of precision and recall.

As shown in Table 1, MSN-DDI achieve the best performance of all six metrics compared with SoTA algorithms on both DrugBank and Twosides. Specifically, although DDI prediction on DrugBank by baseline models is highly accurate, our model still made further improvement on all evaluation metrics. As a comparison, MSN-DDI achieved significant improvement on Twosides by making 19.32% and 17.54% relative improvements on ACC and F1 score over the second-best approach, GMPNN-CS. Furthermore, the AUC reaches 99.47% on DrugBank and 99.90% on Twosides respectively, which indicates MSN-DDI made perfect DDI prediction on transductive setting and basically has solved the DDI prediction task for existing drugs. These results have verified the powerful representational ability of MSN-DDI with several novel-designed components for multi-view substructure learning.

## Performance evaluation on inductive Setting for unseen drugs

In the inductive setting, it is more challenging than the transductive setting since there exists unseen drugs in DDI triplets in test sets. This $cold\text{-}start$ scenario is an extremely difficult trial for the generalization ability of the model, without knowing any prior knowledge of unseen drugs in the training process. In this setting, we split the dataset with respect to the drugs following the common definitions in [19, 20, 26]. Specifically, we randomly picked 20% of drugs as unknown drugs and regarded the remaining drugs as existing ones. All positive and negative samples on the train dataset are all DDI triplets in which both drugs are existing drugs while the test set has two splitting strategies:

- **S1 Partition**: the positive and negative samples on the test set have two unknown drugs. This task is to predict DDI for a pair of new drugs that no effect is known in any combination with other drugs in the training set.
- **S2 Partition**: the positive and negative samples on the test set have one unknown drug and one existing drug. This task is to predict DDI for a new drug that has no effect in any combination with another existing drug.

Furthermore, to avoid potential bias in unknown drug selection, we repeated this process three times in parallel and thus made 3-fold cross validation for the inductive setting.

**Table 2** Performance evaluation of MSN-DDI and baselines for inductive setting on DrugBank dataset (%). The highest value in each column is shown in bold. For performance improvement over the second-best approach, a relative improvement percentage is shown in the bracket.

| Method | S1 Partition (new drug, new drug) | | | | S2 Partition (new drug, existing drug) | | | |
|---|---|---|---|---|---|---|---|---|
| | ACC(%) | AUC(%) | AP(%) | F1(%) | ACC(%) | AUC(%) | AP(%) | F1(%) |
| MR-GNN | 62.63±0.77 | 70.92±0.84 | 73.01±1.23 | 45.81±2.51 | 74.67±0.33 | 83.15±0.60 | 83.81±0.69 | 69.88±0.86 |
| MHCADDI | 66.50±0.62 | 72.53±0.92 | 71.06±1.61 | 67.21±0.59 | 70.58±0.94 | 77.84±1.08 | 76.16±1.45 | 72.74±0.65 |
| SSI-DDI | 65.40±1.30 | 73.43±1.81 | 75.03±1.42 | 54.12±3.46 | 76.38±0.92 | 84.23±1.05 | 84.94±0.76 | 73.54±1.50 |
| GAT-DDI | 66.31±0.61 | 72.75±0.78 | 71.61±1.00 | 68.68±0.60 | 69.83±1.41 | 77.29±1.63 | 75.79±1.95 | 73.01±0.85 |
| GMPNN-CS | 68.57±0.30 | 74.96±0.40 | 75.44±0.50 | 65.32±0.23 | 77.72±0.30 | 84.84±0.15 | 84.87±0.40 | 78.29±0.16 |
| MSN-DDI | **73.42±1.29** | **81.79±1.12** | **81.82±1.48** | **70.34±0.98** | **81.92±1.20** | **91.01±0.76** | **91.09±0.93** | **80.18±1.49** |
| Improvement | +4.85(7.07%) | +6.83(9.11%) | +6.38(8.46%) | +1.66(2.42%) | +4.2(5.4%) | +6.17(7.27%) | +6.15(7.24%) | +1.89(2.41%) |

Previous studies prove that the chemical structures of new drugs in the test set are very different from existing drugs in the training set due to the large differences on scaffolds [19, 20]. As shown in Table 2, all metrics are obvious lower than those evaluated in the transductive setting, which indicates accurate

DDI prediction for unseen drugs is much more difficult. Similar to the performance on the transductive setting, MSN-DDI achieves the best performance on all metrics when compared with state-of-the-art algorithms. Furthermore, our model outperforms the second-best algorithm with a large margin, e.g., relative improvements of 9.11% and 7.27% of AUC on S1 and S2 partitions respectively. These results indicate the effective countermeasures of our model that it does not only consider the intra-structure of the drug itself, but also learns the generalization properties through the multi-view substructure learning framework, which greatly supplements the lack of prior knowledge and interactive information of unseen drugs.

## Ablation study for the effectiveness of model design

To study where the performance gains come from, we perform detailed ablation studies to stress the importance of various components of MSN-DDI. Specifically, the following baselines are evaluated and compared with MSN-DDI:

- **wo_inter**: an architecture where the inter-view message passing from drug-drug interaction module is removed and solely uses the internal message passing. The side effect probability is then computed by concatenating the individual drug representations. This serves to demonstrate the importance of jointly learning drug embeddings (i.e., inter-view interactions).

- **wo_intra**: an architecture where the intra-view message passing in drug itself is removed and solely uses external message passing. The side effect probability is then computed by concatenating the individual drug representations. Similar to above, this serves to demonstrate the importance of simultaneously performing both inter-view and intra-view feature extraction.

- **wo_update**: an architecture where the inter-view interactions are only considered in the substructure extraction module while it has no direct influence on node feature update. This serves to demonstrate the effect of inter-view interaction to drug representation learning.

- **wo_SAGPool**: an architecture where the readout function in the substructure extraction module is replaced by a simple sum function, without distinguishing the importance of nodes. In this setting, both the inter and intra drug embeddings are computed by the sum readout function. This serves to demonstrate the importance of performing self-attention graph pooling for substructure extraction.

- **wo_co-attention**: an architecture where the final DDI prediction is also computed from the pairwise interactions between substructures of a pair of drugs but without using the global attention among all substructures extracted from MSN encoder blocks as shown in Eq. **??**. As it has been also used in previous studies [19], this serves to evaluate ist contribution to our performance gains.

**Table 3** Performance evaluation between MSN-DDI and its five variants on DrugBank dataset in inductive setting. The highest value in each column is shown in bold.

| Method | S1 Partition | | | | S2 Partition | | | |
|---|---|---|---|---|---|---|---|---|
| | ACC(%) | AUC(%) | AP(%) | F1(%) | ACC(%) | AUC(%) | AP(%) | F1(%) |
| wo_inter | 65.45 | 72.67 | 73.69 | 57.94 | 75.30 | 83.08 | 84.69 | 71.51 |
| wo_intra | 68.23 | 75.81 | 76.07 | 64.06 | 76.94 | 85.93 | 86.16 | 74.13 |
| wo_update | 66.52 | 74.05 | 75.40 | 58.71 | 75.57 | 84.13 | 85.13 | 71.74 |
| wo_SAGPool | 69.71 | 77.05 | 77.03 | 66.11 | 79.31 | 87.76 | 87.62 | 77.77 |
| wo_co-attention | 71.36 | 78.56 | 77.14 | 69.98 | 78.65 | 86.69 | 86.63 | 76.94 |
| MSN-DDI | **73.42** | **81.79** | **81.82** | **70.34** | **81.92** | **91.01** | **91.09** | **80.18** |

As shown in Table S3 and Table 3, the full MSN-DDI architecture outperformed all variants which indicates the effectiveness of the proposed methods. In particular, our method significantly outperforms all other variants on the inductive setting, showing considerable modeling advantages over MSN-DDI. We further summarize the conclusions as follows: (1) Inter-view contributes most to MSN-DDI, since the performance of the variant wo_inter decreases significantly. The fact that the performance of two variants wo_inter and wo_intra decline to some extent implies that it is beneficial to learn drug-drug representations jointly from multi-view perspective rather than a respective view. (2) When comparing between the variant wo_update and MSN-DDI, the performance has also declined with a large margin, which indicates that besides for substructure extraction and final DDI prediction, the inter-view information is useful for drug representation learning when directly incorporated into node update process. (3) As we can from the results of wo_SAGPool in inductive settings on Twosides datasets, the performance of the model without the final SAGPool module does not decline significantly, which reflects that the improvement brought by our model in this task does not depend on the specific readout function. (4) The little performance drop on wo_co-attention indicates that the co-attention mechanism only plays a minor role in model performance gains, which reflects substructures are distinct and robust and can directly be used for DDI prediction without such complicated attention mechanism. As can be seen from the above results, the two new modules proposed in MSN-DDI have greatly improved the performance of DDI prediction on Twosides dataset, including multi-view interaction and update  module.

## Real-world DDI applications

To demonstrate the usefulness of MSN-DDI for real-world DDI applications, we first evaluate the DDI prediction for new FDA approved drugs by the model trained with existing information of old drugs. We collected the FDA drug approval information [27] for all drugs in DrguBank dataset and divided them into two parts according to the drug approval date before or after the year of 2017. The DDI triplets containing two old drugs form the training set while the remaining DDI triplets containing at least one new drug are recruited into the test set (see Supplementary Table S5 for more details). We trained and evaluated MSN-DDI with the same hyperparameters adopted in the inductive setting. Furthermore, we also picked the two state-of-the-art DDI prediction algorithms from the above performance evaluation, SSI-DDI and GMPNNN-CS for comparison. These two algorithms were reproduced on the same dataset with their default hyperparameters. As shown in Supplementary Table S6, MSN-DDI outperformed SSI-DDI and GMPNN-CS on all four metrics, ACC, AUROC, AP and F1 with a large margin. These results consolidate MSN-DDI has captured the generalized information of drug-drug interaction among different drugs and thus is applicable for new approved drugs. Figure 2 illustrated the detailed ROC curve and PR curve on the test set of three algorithms. MSN-DDI achieved the significant larger areas under both the ROC and PR curves than SSI-DDI and GMPNN-CS respectively, which also indicates MSN-DDI can distinguish the positive DDI effects from negative ones well.

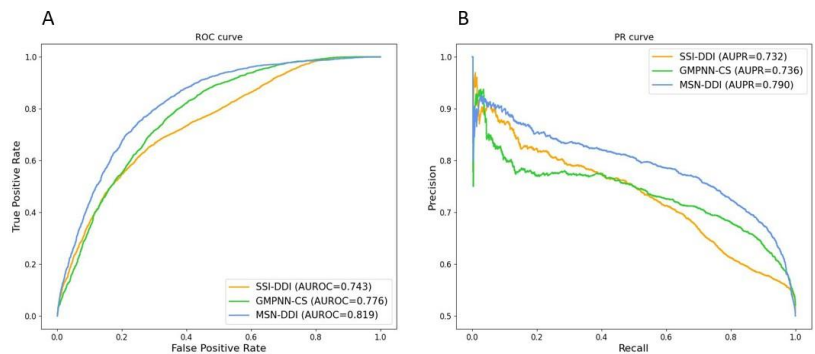

**Fig. 2** Performance evaluation of SSI-DDI, GMPNN-CS and MSN-DDI for new approved drugs. (A) The receiver operating characteristic (ROC) curve of three algorithms evaluated on the test set. (B) The prevision versus recall (PR) curve of three algorithms evaluated on the test set.

Furthermore, we exhibit the usefulness of MSN-DDI by a case study of drug combination for anti-COVID-19. We utilize MSN-DDI model trained on the DrugBank dataset to predict the probability of the triplet *(Hydroxychloroquine, increase the QTc prolongation, Azithromycin)*, where the two drugs *Hydroxychloroquine* and *Azithromycin* are both known drugs for our model, which can be seen as the transductive setting. These two drugs were recommended as a combination of potential anti COVID-19 drugs when an outbreak of the pandemic, but were found to have the serious side effect. It has been verified that the risk or severity of QTc prolongation can be increased when *Hydroxychloroquine* is combined with *Azithromycin* [28], and our model can effectively filter such drug combinations and thus contribute to therapies for anti-COVID-19. It has been verified that the risk or severity of QTc prolongation can be increased when *Hydroxychloroquine* is combined with *Azithromycin* [28].

In Figure 3, we extract and illustrate the valid substructures of the two drugs in the event of increasing the QTc prolongation from inter-view, intra-view and multi-view, respectively. Specifically, as demonstrated in the ablation study, intra-view and inter-view are two variants of our model, which remove the inter-view interaction and intra-view interaction, respectively while the multi-view panel is the full MSN-DDI neural network. We first used the SAG-pooling layers to obtain the contribution scores of each atom in drugs, and extract the top k atoms as important atoms based on contribution scores for each block *(k = 10 for Hydroxychloroquine and k = 15 for Azithromycin)*. Then, we count and select the important atoms with more than 3 occurrences in all blocks as the final learned substructures and highlighted them with shadows colored in green (Figure 3).

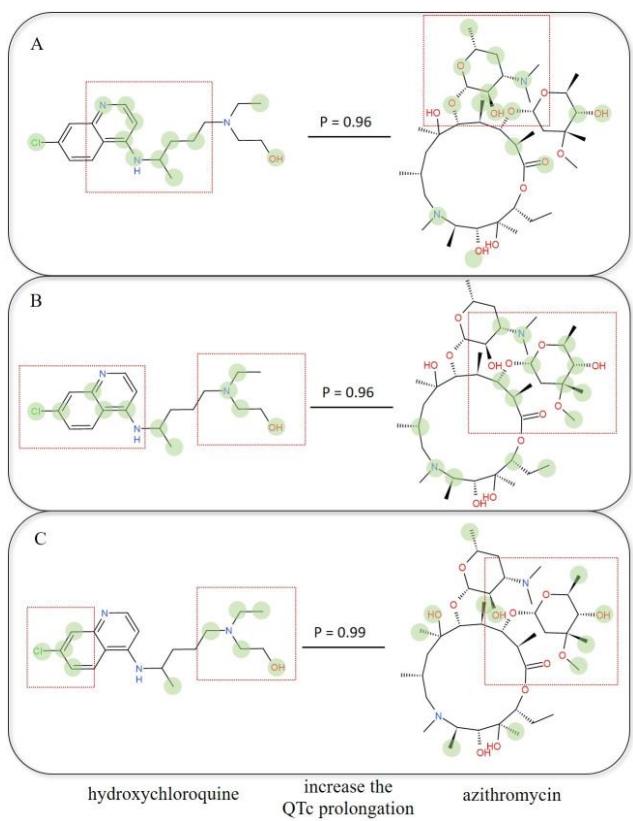

**Fig. 3** Visualization of DDI prediction on the triplet *(Hydroxychloroquine, increase the QTc prolongation, Azithromycin)* from intra-view (A), inter-view (B) and multi-view (C), respectively. The learned substructures are made up with the atoms highlighted with shadows colored in green, which are selected in terms of high contribution scores in DDI prediction and high occurrences in all blocks in our model.

As shown in Figure 3 (A) and (B), the highlighted atoms learned from single perspective (intra-view or inter-view) are significantly different, dispersed in the whole structure of the drug and fail to form certain substructures, which may lead to lower DDI prediction values. As a comparison, important atoms learned by MSN-DDI are centralized to some certain regions of the drug chemical structure to form steady substructures and these atoms can be regarded as stacking the important atoms from both the inter-view and intra-view. As a result. MSN-DDI with multi-view made a perfect prediction (i.e., a prediction score of 0.99) for this case. This analysis implies that MSN-DDI is not only a good DDI predictor but also could provide some clues to interpret the DDI effect to possible interactive atoms, which may facilitate detecting the underlying mechanism of the combination of drug pairs.

## Conclusions

In this work, we have presented a multi-view substructure learning framework for predicting the possible polypharmacy side effects of drug-drug combination. Extensive experiments have verified the state-of-the-art performance for DDI prediction on both transductive and inductive settings. The MSN-DDI has achieved significant improvements with 19.32% accuracy on Twosides in transductive setting compared with the SOTA methods. More importantly, the performance of our proposed method has achieved significant improvement in more challenging inductive scenarios, with an average improvement 7.07% on DrugBank and 5.40% on Twosides in accuracy compared with the SoTA methods. By performing intra-view message passing within each drug, as well as inter-view message passing between two drugs, we have demonstrated the power of integrating joint drug-drug information during the substructure representation learning phase for DDI prediction. Future directions could put more attention on the generalization of the model for new drugs in the inductive learning setting, which approximates a real-world scenario where there is a new drug without knowing prior associated drug interactions.